\documentclass[useAMS,usenatbib]{mn2e}

\usepackage{graphicx}
\usepackage{color}

\def\gsim{\mathrel{\rlap{\lower 4pt \hbox{\hskip 1pt $\sim$}}\raise 1pt
\hbox {$>$}}}
\def\lsim{\mathrel{\rlap{\lower 4pt \hbox{\hskip 1pt $\sim$}}\raise 1pt
\hbox {$<$}}}

\title[CS Dust around SNe Ia]{Constraining the Amount of Circumstellar Matter and Dust around Type Ia Supernovae through Near-Infrared Echoes}
\author[K. Maeda et al.]{Keiichi Maeda$^{1,2}$\thanks{E-mail: keiichi.maeda@kusastro.kyoto-u.ac.jp}, Takaya Nozawa$^{3}$, Takashi Nagao$^{1}$, Kentaro Motohara$^{4}$\\
$^{1}$Department of Astronomy, Kyoto University, 
Kitashirakawa-Oiwake-cho, Sakyo-ku, Kyoto 606-8502, Japan\\
$^{2}$Kavli Institute for the Physics and Mathematics of the 
Universe (WPI), University of Tokyo, \\
5-1-5 Kashiwanoha, Kashiwa, Chiba 277-8583, Japan\\
$^{3}$National Astronomical Observatory of Japan, 2-21-1 Osawa, Mitaka, Tokyo, 188-8588, Japan\\
$^{4}$Institute of Astronomy, Graduate School of Science, University of Tokyo, 2-21-1 Osawa, Mitaka, Tokyo 181-0015, Japan}
\begin{document}

\date{}

\pagerange{\pageref{firstpage}--\pageref{lastpage}} \pubyear{2014}

\maketitle

\label{firstpage}

\begin{abstract}
The circumstellar (CS) environment is key to understanding progenitors of type Ia supernovae (SNe Ia), as well as the origin of a peculiar extinction property toward SNe Ia for cosmological application. It has been suggested that multiple {\em scatterings} of SN photons by CS dust may explain the non-standard reddening law. In this paper, we examine the effect of {\em re-emission} of SN photons by CS dust in the infrared (IR) wavelength regime. This effect allows the observed IR light curves to be used as a constraint on the position/size and the amount of CS dust. The method was applied to observed near-infrared (NIR) SN Ia samples; meaningful upper limits on the CS dust mass were derived even under conservative assumptions. We thereby clarify a difficulty associated with the CS dust scattering model as a general explanation for the peculiar reddening law, while it may still apply to a sub-sample of highly reddened SNe Ia. For SNe Ia in general, the environment at the interstellar scale appears to be responsible for the non-standard extinction law. Furthermore, deeper limits can be obtained using the standard nature of SN Ia NIR light curves. In this application, an upper limit of $\dot M \lsim 10^{-8} - 10^{-7} M_{\odot}$ yr$^{-1}$ (for the wind velocity of $\sim 10$ km s$^{-1}$) is obtained for a mass loss rate from a progenitor up to $\sim 0.01$ pc, and $\dot M \lsim 10^{-7} - 10^{-6} M_{\odot}$ yr$^{-1}$ up to $\sim 0.1$ pc. 
\end{abstract}

\begin{keywords}
Circumstellar matter -- 
stars: mass-loss -- 
dust, extinction -- 
supernovae: general. 
\end{keywords}

\section{Introduction}

Type Ia supernovae (SNe Ia) are mature standardised candles. Intrinsic dispersion of the peak absolute magnitude can be minimised to the level of $\sim 0.15$ magnitude or even smaller, and can be accurately used as a cosmological distance indicator \citep{phillips1999}. However, this standardisation generally requires a non-standard extinction law, i.e., $R_V \equiv A_V / E (B-V)\lsim 2$, as opposed to the typical galactic value of $R_V \sim 3.1$ \citep[e.g., ][]{folatelli2010}. This indicates that the properties of dust in SN Ia host galaxies may be systematically different from those in our galaxy \citep[e.g.,][]{phillips2013}. 

Alternatively, it has been suggested that multiple {\em scatterings} of SN photons on circumstellar (CS) dust may explain the non-standard nature of the reddening law \citep{wang2005,goobar2008}. Despite the implications that have been discussed \citep[e.g.,][]{folatelli2010}, researchers have yet to find a smoking gun to test this hypothesis, apart from the various model predictions for optical wavelengths \citep{amanullah2011}. This interpretation is also associated with as yet unclarified progenitor evolution, through the dense circumstellar medium (CSM) and high mass loss rate that would be required. Qualitatively, the scenario favours a so-called single-degenerate model \citep[i.e., a pair consisting of a C+O white dwarf and a non-degenerate star; ][]{whelan1973,nomoto1982,hachisu1999}, which should create a relatively dense (or `dirty') CS environment. 

In this paper, we examine the effects of {\em re-emission} of SN photons for infrared (IR) wavelengths. While the {\em `scattering'} must be accompanied by {\em `absorption and re-emission'} (a CS dust echo), this process has not been examined with respect to the CS environment around SNe Ia \citep[with the exception of recent work that focused on the mid- and far-IR; ][]{johansson2013}. The remainder of this paper is organised as follows. In \S 2, we develop a simple echo model for CS dust heated by SN photons. In \S 3, we apply the method to observed NIR SN Ia samples. We place both conservative and deeper limits for the amount of CS dust under different assumptions. In \S 4, we discuss the effects of different types of CS dust species, as well as future observational strategies for a better constraint on the amount of CS dust and the environment. The paper is closed in \S 5 with conclusions. 

\section{Models}

In this section, we present our model of CS dust echoes. We restrict the model to absorption and re-emission processes, not including scattering. This is justified because, (1) given that the spectral peak of the SN light lies at optical wavelengths, the contribution of the scattered SN light at NIR-IR wavelengths is very small, compared to that of thermal emission from CS dust, and (2) in the distribution of CS dust of interest in this paper, further scattering of the re-emitted NIR/IR photons is negligible. While computational methods for an echo from spherically distributed CS dust \citep[e.g., ][]{dwek1983,dwek1985} and further from aspherically distributed CS dust \citep[especially for modelling the light echo from SN 1987A; ][]{dwek1989,felten1989} have been developed, we assume for the sake of simplicity that the distribution of CS dust is represented by an infinitely thin shell located at radius $R$ from the centre of an explosion. The thin-shell approximation further simplifies the computation as shown below. Furthermore, in providing an upper limit for the amount of CS dust {\em as a function of} $R$, the thin-shell approximation basically provides the most conservative limit, as any spherically symmetric distribution can be described as a convolution of different shells. Namely, if one adds another shell at $R^{\prime}$, that is different from $R$ for which the constraint is considered, then this additional shell will increase the predicted flux without contributing to the CS dust mass at $R$; therefore, the upper limit on this CS dust mass should be reduced (i.e., the predicted flux must be below observations). 

\begin{figure*}
\centering
 \includegraphics[width=0.45\textwidth]{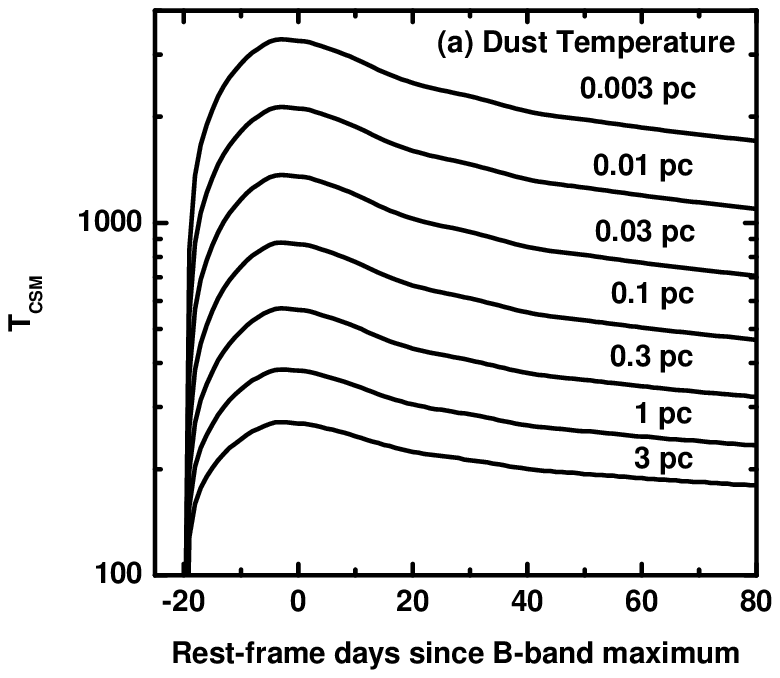}
 \includegraphics[width=0.45\textwidth]{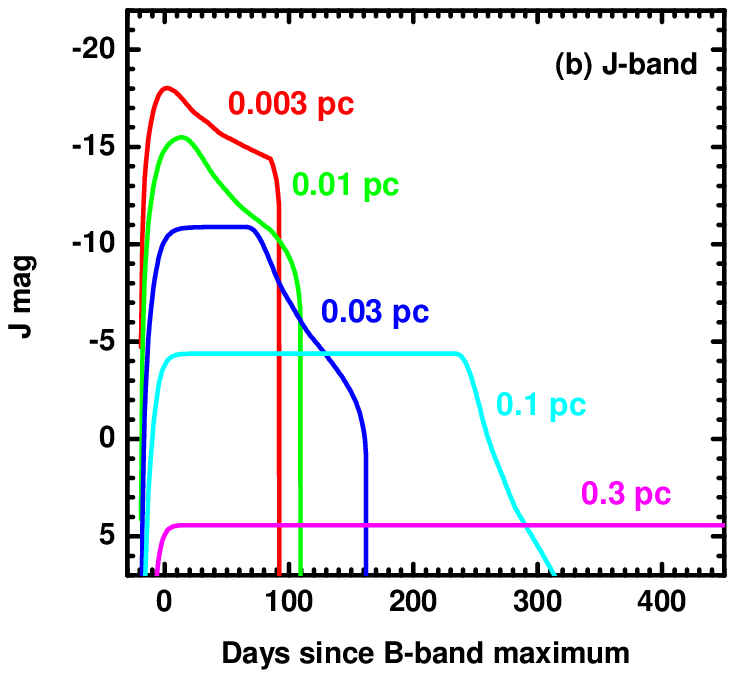}
 \includegraphics[width=0.45\textwidth]{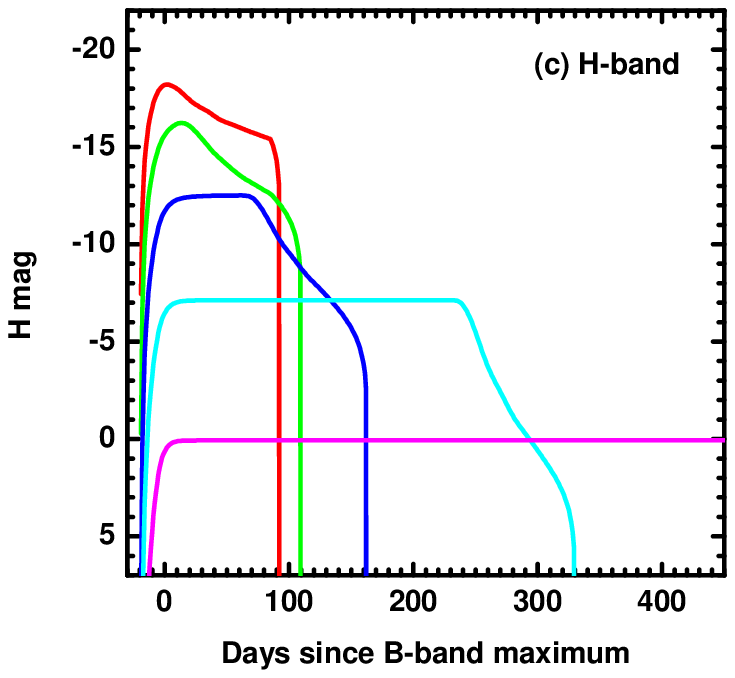}
 \includegraphics[width=0.45\textwidth]{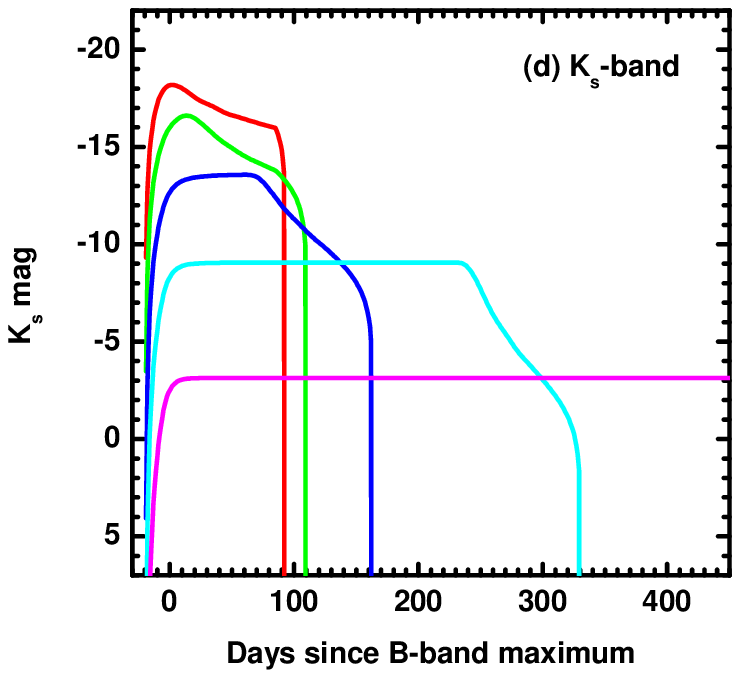}
\caption
{NIR light echo models. (a) Temperature of the CS dust ($T$) as a function of the {\em rest-frame} days, in which the $B$-band maximum is shown for various distances to the CS dust ($R = 3.24 \times 10^{-3}$, $1.02 \times 10^{-2}$, $3.24 \times 10^{-2}$, $1.02 \times 10^{-1}$, $3.2 \times 10^{-1}$, $1.02$, and $3.24$ pc from top to bottom). In all of the models, the CS dust mass is set to $10^{-5} M_{\odot}$. (b--d) The echo model light curves for $J$ (b), $H$ (c), and $K_{\rm s}$-bands (d) as a function of the {\em observed} days for the shell model. The colour coordinate denotes the model for different distances to the CS dust (red for $R = 3.24 \times 10^{-3}$ pc, green for $1.02 \times 10^{-2}$ pc, blue for $3.24 \times 10^{-2}$ pc, cyan for $1.02 \times 10^{-1}$ pc, and magenta for $3.24 \times 10^{-1}$ pc).}
\label{fig1}
\end{figure*}

\begin{figure*}
\centering
 \includegraphics[width=0.33\textwidth]{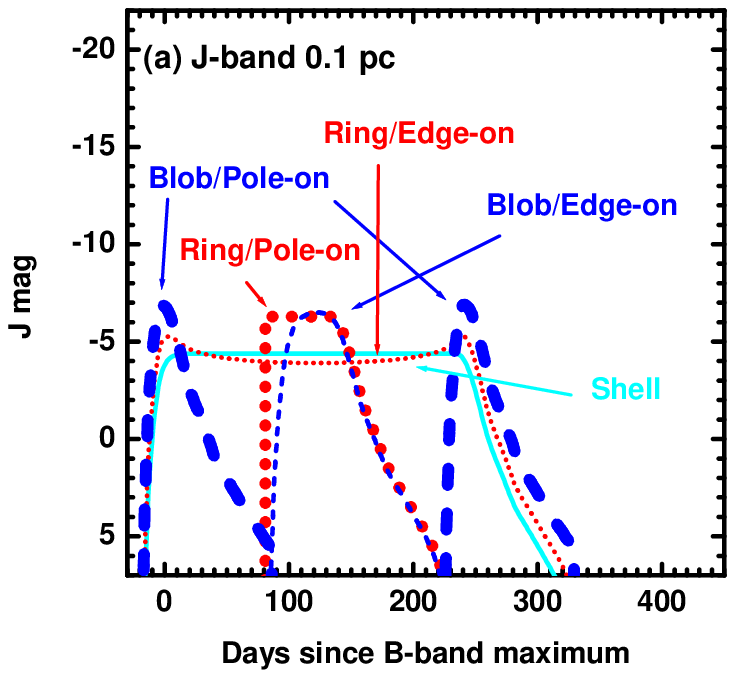}
 \includegraphics[width=0.33\textwidth]{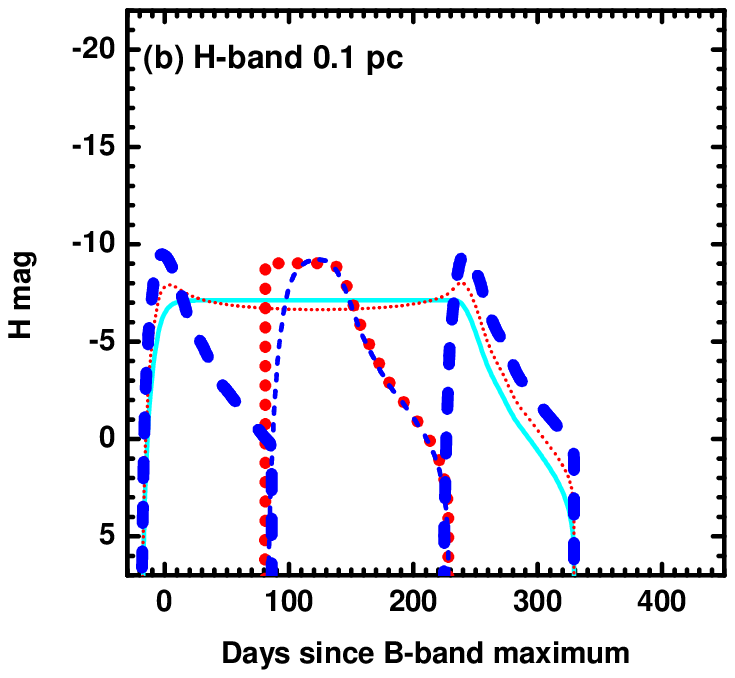}
 \includegraphics[width=0.33\textwidth]{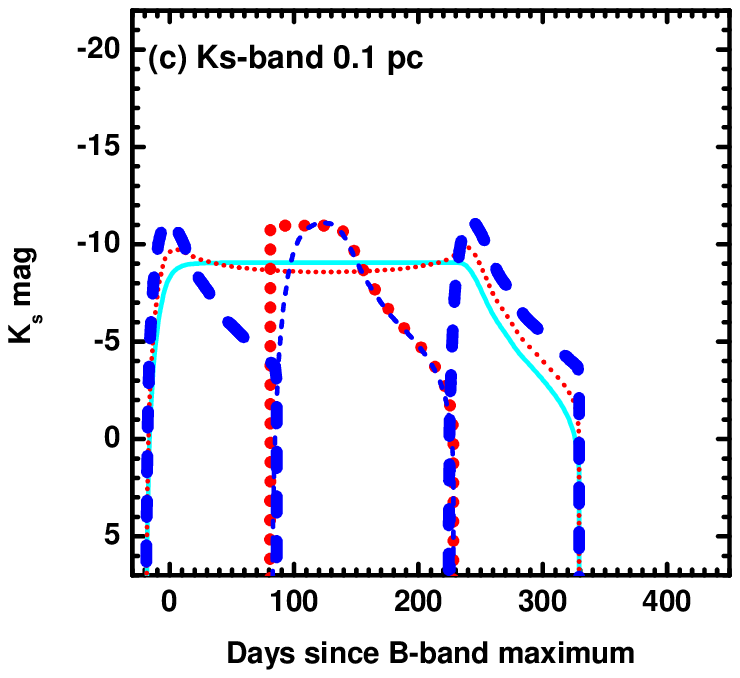}
\caption
{Effect of different geometries on the echo light curves. For demonstration purposes, the distance to the CS dust is fixed at $R = 0.1$ pc. The reference shell model is shown in cyan (solid). The echo light curve from a ring-like CS dust distribution is shown by red (dotted) curves for an edge-on observer (thin) and a pole-on observer (thick). The same is shown for a blob-like CS dust distribution by blue (dashed) curves for an edge-on observer (thin) and for a pole-on observer (thick). 
}
\label{fig2}
\end{figure*}

\subsection{Shell Model} 

Let us first consider a situation in which we observe a dust shell with temperature $T (\theta)$ and mass $M_{\rm d}$, where $\theta$ is the angle between the observer's line of sight and a direction vector pointing from the shell centre (i.e., the explosion centre) to the position of the volume element within the shell under consideration. Then, the luminosity of the thermal emission from dust is given as follows: 

\begin{eqnarray}
L_{{\rm echo}, \nu} & = & \int_{V} 4 \pi \kappa_{{\rm a}, \nu} B_{\nu} (T (\theta)) \rho dV \nonumber\\
& = & 2 \pi M_{\rm d} \kappa_{{\rm a}, \nu} \int_{0}^{\pi} \sin{\theta} B_{\nu} (T(\theta)) d\theta \ .
\end{eqnarray}
Here, $\kappa_{{\rm a}, \nu}$ is the absorbing opacity of the dust at frequency $\nu$. It is specified by the dust properties, which are assumed to be uniform within the shell. Furthermore, the shell is assumed to have uniform density ($\rho$). The above expression can be converted into time-dependent luminosity by introducing $t$ (the time since the explosion, measured in the observer's frame) and $t^{\prime}$ (observer-frame time when the light was emitted from a given volume element at $\theta$ as observed at time $t$). The relationship between $t$ and $t^{\prime}$ is given by the following: 
\begin{equation}
t^{\prime} = t - \frac{R}{c} (1 - \cos{\theta}) \ .
\end{equation}
Combining Eqs. (1) and (2), we obtain the following expression for the echo luminosity at time $t$: 
\begin{equation}
L_{{\rm echo}, \nu} (t) = 2 \pi \frac{c}{R} M_{\rm d} \kappa_{{\rm a}, \nu} \int_{{\rm max} (t-\frac{2R}{c}, 0)}^{t} B_{\nu} (T(t^{\prime})) dt^{\prime} \ . 
\end{equation}

The temperature evolution of the CS dust at time $t$, under radiative equilibrium with an incoming SN flux, $L_{{\rm SN}, \nu} (t)$, is given by 
\begin{equation}
\int_{0}^{\infty} \frac{L_{{\rm SN}, \nu} (t)}{4 \pi R^2} \kappa_{{\rm a}, \nu} d\nu = 4 \pi \int_{0}^{\infty} \kappa_{{\rm a}, \nu} B_{\nu} (T (t)) d\nu \ . 
\end{equation}
This equation provides $T (t)$ once the SN flux evolution is specified. Under the condition considered in this paper, the photon-dust collision time scale is shorter than the cooling time scale of the dust by thermal radiation. Therefore, stochastic heating is negligible, and we can assume that virtually all of the dust particles have a temperature determined by radiative equilibrium.

We first compute $T (t)$ for a given $R$ using the above equation. For the SN flux, we use the Hsiao template spectra as an input \citep{hsiao2007}. For the dust opacity, we use the LMC-like dust properties as our fiducial model; specifically, we adopt the opacity from $U$ to $K_{\rm s}$ from \citet{goobar2008} so as to be consistent with the proposed CS scattering model. We extend the opacity toward the blue (beyond $U$) and red (beyond $K$), assuming slopes of $-1.5$ and $-1$, respectively. Then the evolution of $T (t)$ is used as an input in the computations of echo luminosity and multi-band light curves. 

Because the incoming SN light peaks in optical wavelengths, and because the dust temperature under the situation examined in this paper is determined by the opacity in the NIR wavelength, the treatment of the opacity below the $U$-band and above the $K_{\rm s}$-band does not substantially affect our results. To check this, we performed two additional test calculations: one having a slope of $-2.0$ below the $U$-band and the other having a slope of $-0.5$ above the $K$-band for our reference LMC dust model; these models are unrealistically steep and flat, respectively, and thus provide a conservative measure of the uncertainty for the treatment of different slopes. The difference between the reference model and the one using the different slopes ($dJ$, $dH$, $dK_{\rm s}$) is larger for a model with a larger $R$. For the typical maximum distance for which a meaningful constraint is placed at different bands (see \S 3), the difference in the echo magnitude is given by $dJ = -0.12$ mag (for 0.03 pc), $dH = -0.30$ mag (for 0.3 pc), and $dK_{\rm s} = -0.31$ mag (1 pc) for the different slope below the $U$-band. For the different slope above the $K_{\rm s}$-band, $dJ = 0.2$ mag (for 0.03 pc), $dH = 1.4$ mag (for 0.3 pc), and $dK_{\rm s} = 1.9$ mag (for 1 pc). Therefore, in the worst-case scenario, our echo luminosity constraint (and, thus, the dust mass) could have an uncertainty factor of 5; however, in most cases the uncertainty would be much smaller than this. A further discussion of the dust properties is given in \S 4.1. 

The resulting echo luminosity, as computed above, is for unobscured luminosity. We now consider the optical depth within the shell. The average optical depth is computed as $\tau_{\nu} = \kappa_{{\rm a}, \nu} \rho \Delta R$; then the NIR echo luminosity is dimmed by this amount of absorption. The same procedure is adopted for the model with a different geometry (see below).

\subsection{Ring/torus model}

The CSM around an SN Ia may be confined within the equatorial plane \citep[e.g.,][]{dilday2012}, therefore we also consider a torus/ring-like structure, again under the infinitely thin assumption in the radial direction. We denote $\theta_0$ as the opening angle of the ring, as measured from the equatorial direction. While it is possible to derive general expressions for an arbitrary viewing direction, hereafter, we focus on two extreme cases: pole-on and edge-on viewing. 

For an observer sitting at the pole-on position, the echo luminosity is expressed as follows: 
\begin{equation}
L_{\rm echo} (t) = 2 \pi \frac{c}{R} \frac{M_{\rm d}}{\sin{\theta_0}} \kappa_{{\rm a}, \nu} \int_{{\rm max} (t-\frac{R}{c} (1 + \sin{\theta_0}), 0)}^{t-\frac{R}{c} (1 - \sin{\theta_0})} B_{\nu} (T(t^{\prime})) dt^{\prime} \ .
\end{equation}
If $t - \frac{R}{c} (1+\sin{\theta_0}) < 0$, then the integral is set to zero. 

To be an explanation for the non-standard extinction law requires $R \lsim c t_{\rm peak} \sim 5 \times 10^{16}$ cm, to introduce the effect of scattered photons around the maximum light. An observer in the equatorial direction does not have this constraint. We note that both situations should be considered simultaneously, because there should be roughly an equal number of edge-on and pole-on counterparts (depending on $\theta_0$), if we adopt the ring-like geometry. Finally, the echo luminosity arising from the same configuration, but viewed edge-on, is given as follows. 
\begin{eqnarray}
& L_{{\rm echo}, \nu} (t) & = 2 \pi \frac{c}{R} \frac{M_{\rm d}}{\sin{\theta_0}} \kappa_{{\rm a}, \nu} [ \int_{{\rm max} (t-\frac{2R}{c}, 0)}^{t-\frac{R}{c} (1 + \cos{\theta_0})} B_{\nu} (T(t^{\prime})) dt^{\prime} \nonumber\\
& + & \int_{t-\frac{R}{c} (1+\cos{\theta_0})}^{t-\frac{R}{c} (1-\cos{\theta_0})} \frac{2}{\pi} B_{\nu} (T(t^{\prime})) \sin^{-1}\left(\frac{\sin{\theta_0}}{\sin{\theta}}\right) dt^{\prime} \nonumber\\ 
& + & \int_{t-\frac{R}{c} (1-\cos{\theta_0})}^{t} B_{\nu} (T(t^{\prime})) dt^{\prime} ] \ .
\end{eqnarray}

\subsection{Bipolar blob model}
Another interesting possibility for the CS dust distribution is a bipolar morphology, i.e., a pair of blobs/jets, as inferred for some novae \citep[e.g., ][]{chesneau2012}. Therefore, we also consider a pair of bipolar blobs for the distribution of the CS dust, again under the infinitely thin assumption in the radial direction. We denote $\theta_0$ as the opening angle subtended by each blob, as measured from the polar direction. 

For an observer sitting at the pole-on position, the echo luminosity is expressed as follows: 
\begin{eqnarray}
& L_{{\rm echo}, \nu} (t) & = 2 \pi \frac{c}{R} \frac{M_{\rm d}}{1-\cos{\theta_0}} \kappa_{{\rm a}, \nu} [ \int_{{\rm max} (t-\frac{2R}{c}, 0)}^{t-\frac{R}{c} (1 + \cos{\theta_0})} B_{\nu} (T(t^{\prime})) dt^{\prime} \nonumber\\
& + & \int_{t-\frac{R}{c} (1-\cos{\theta_0})}^{t} B_{\nu} (T(t^{\prime})) dt^{\prime} ] \ .
\end{eqnarray}
The echo luminosity arising from the same configuration, but viewed edge-on, is given as 
\begin{eqnarray}
& L_{{\rm echo}, \nu} (t) & = 2 \pi \frac{c}{R} \frac{M_{\rm d}}{1-\cos{\theta_0}} \kappa_{{\rm a}, \nu} \int_{t-\frac{R}{c} (1+\sin{\theta_0})}^{t-\frac{R}{c} (1-\sin{\theta_0})} \\
& \times & \frac{2}{\pi} B_{\nu} (T(t^{\prime})) \cos^{-1}\left(\frac{\cos{\theta_0}}{\sin{\theta}}\right) dt^{\prime} \ .
\end{eqnarray}
Similar to the case of the pole-on observer for the ring-like CS dust distribution, a constraint is placed on the CS dust radius if the multiple scattering model is to account for the non-standard extinction law. For the edge-on observer of the bipolar blob geometry, the CS dust radius must satisfy $R \lsim c t_{\rm peak} \sim 5 \times 10^{16}$ cm to introduce the scattered photon effect around the maximum light. This constraint does not apply to the pole-on observer; however, both situations (pole-on and edge-on) should be considered simultaneously, following the same argument given for the ring-like geometry. 

\section{Results}

Figure 1 shows the evolution of $T$ and the predicted NIR ($J$, $H$, $K_{\rm s}$) light curves for different values of $R$ and for a fixed value of $M_{\rm d}$ ($10^{-5} M_{\odot}$); for the light curves, only the shell model is shown. The evolution of $T$ closely follows the optical light curve of SN Ia. For $R\lsim 0.01$ pc, the peak temperature exceeds $\sim 2,000$K; therefore, the dust may well have evaporated, depending on the nature of the CS dust. In such a case, essentially there should be neither scattering nor thermal emission from the CS dust. Therefore, no constraint can be obtained regarding the amount of pre-existing CS dust, {\em and} the CS dust cannot be an origin of the non-standard extinction law toward SNe Ia. The effects of different geometries are demonstrated in Fig. 2; light curves are shown for the shell CSM, pole-on and edge-on views for a ring-like CSM with $\theta_0 =10^{\circ}$, and pole-on and edge-on views for bipolar blobs with $\theta_0 = 10^{\circ}$. Throughout this paper, we consider these geometry types for the CS dust distribution, while adopting the shell model as our fiducial model. 

\begin{figure*}
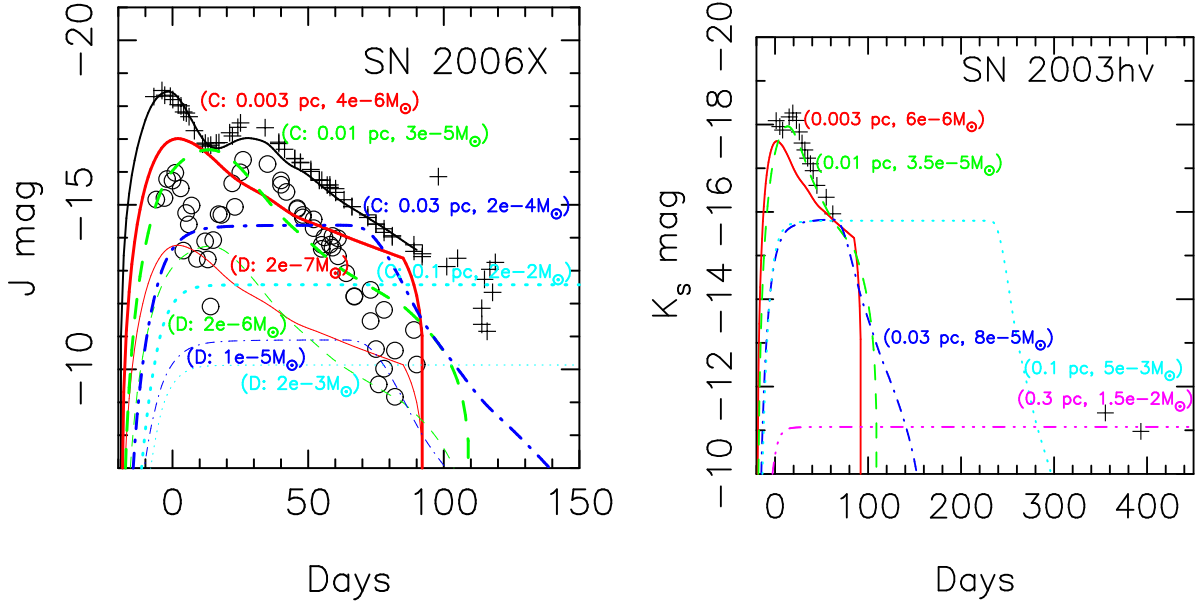

\centering
 \includegraphics[width=0.45\textwidth]{f3a.eps}
 \includegraphics[width=0.45\textwidth]{f3b.eps}
\caption
{Examples of the light curve constraints on the CS dust mass, applied to the $J$-band light curve of SN 2006X \citep[left; ][]{wood2008} (also shown is the Hsiao template, represented by a black solid line) and the $K_{\rm s}$-band light curves of SN 2003hv \citep[right; ][]{motohara2006}. For SN 2006X, two {\em observational} light curves are shown -- one being the original light curve (crosses) and the other constructed from the {\em residual} after subtracting the template light curve (open circles; see main text). The shell CS dust echo modelled with the {\em maximally allowed} CS dust mass are shown for select values of $R$ (red/solid for $3.24 \times 10^{-3}$ pc, green/dashed for $1.02 \times 10^{-2}$ pc, blue/dot-dashed for $3.24 \times 10^{-2}$ pc, cyan/dotted for $1.02 \times 10^{-1}$ pc, and magenta/dots-dashed for $3.24 \times 10^{-1}$ pc). For SN 2006X, two models are shown for each $R$ -- one using the original light curve (thick, `C' in the labels) and the other using the residual light curve (thin, `D' in the labels). These examples highlight the importance of good early-phase light curves to constraint the CS dust mass at small $R$ especially coupled with the template light curves, as well as deep late-time photometric points preferably at longer wavelengths to constrain the CS dust mass at large $R$. 
}
\label{fig3}
\end{figure*}

The light curve evolution is similar in various NIR bands; its time scale is basically determined by the travelling time of light, and thus the CS dust at larger $R$ results in a longer-lived, fainter NIR echo. For small $R$, the travelling time of light is shorter than the time scale of the SN light curve evolution; thus in this case, the light curve reflects the SN light curve evolution, and the echo light curve evolution traces the decreasing temperature of the CS dust as a function of time. On the other hand, for large $R$, the travelling time of light far exceeds the SN light curve evolution time scale; therefore, the temperature evolution has little influence on the shape of the light curve. Emission at longer wavelengths is stronger for larger $R$, which stems from smaller $T$ for larger $R$. As such, the peak wavelength in the thermal emission is shorter than the $J$-band for smaller $R$, while it is even longer than the $K_{\rm s}$-band pass for larger $R$. 

Figure 3 shows examples of the constraint on the amount of CS dust using the shell model. Without the dust echo, NIR luminosities of SNe are powered by radioactive decay of $^{56}$Co and subsequent thermalisation; the NIR spectral energy distribution (SED) roughly follows a blackbody of $\sim 5,000 - 10,000$ K in the early photospheric phase ($\lsim 100$ days), while the SED reflects the forbidden lines of Fe-peak elements in late nebular phases ($\gsim 100$ days). In any case, a conservative upper limit on the CS dust mass is obtained by the condition that the resulting echo luminosity cannot exceed the observation at any epochs. As the echo luminosity is proportional to $M_{\rm d}$ in the optically thin regime, for a given $R$, one can derive a maximally allowed value for $M_{\rm d}$. Note that this is not a fit to the observed light curve -- the echo light curve model for small $R$ resembles the observed light curve to some extent, but this is because the echo evolution follows the SN evolution for small $R$ (see above). 

The left panel of Fig. 3 shows how an early-phase light curve (up to $\sim 100$ days since the $B$-band maximum) can be used to constrain the amount of CS dust. Generally, the early phase data provide a strong constraint on $M_{\rm d}$ for small $R$, but lose diagnostic power for CS dust at large $R$. The model prediction for small $R$ is similar for different bands; typically the $J$-band provides the deepest limit, thanks to its extensive coverage within the NIR. We note that the $I$-band, which typically has much better coverage than the $J$-band, is not so useful for this purpose, as the wavelength of the thermal emission peak is longer than that of the $I$-band. 

The right panel of Fig. 3 shows the other extreme within the NIR wavelengths, namely a constraint obtained through late-time photometry in the $K_{\rm s}$-band. Such an observation can provide a constraint on the CS dust at large $R$, up to $\sim 1$ pc.

\begin{figure*}
\centering
 \includegraphics[width=0.33\textwidth]{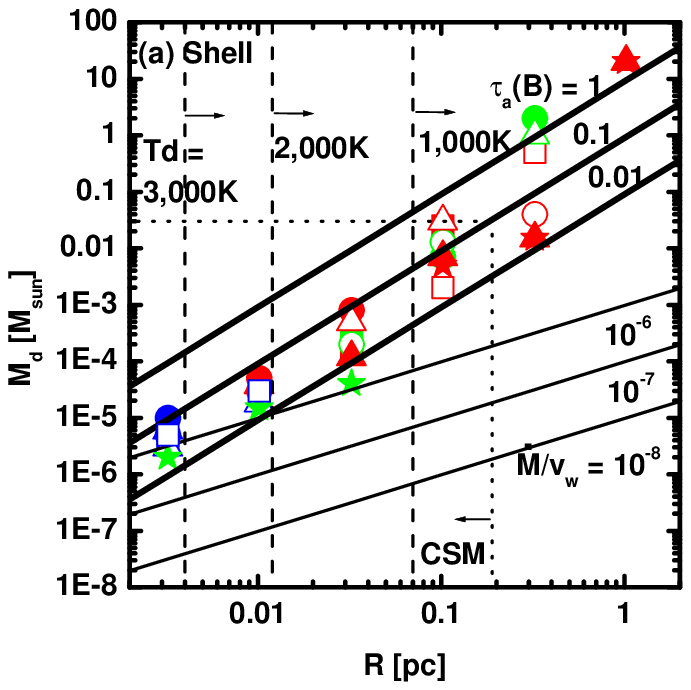}
 \includegraphics[width=0.33\textwidth]{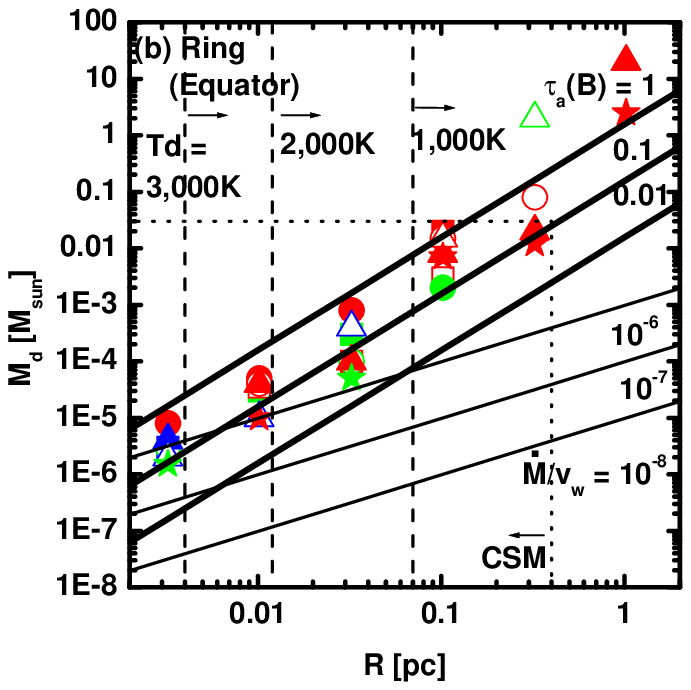}
 \includegraphics[width=0.33\textwidth]{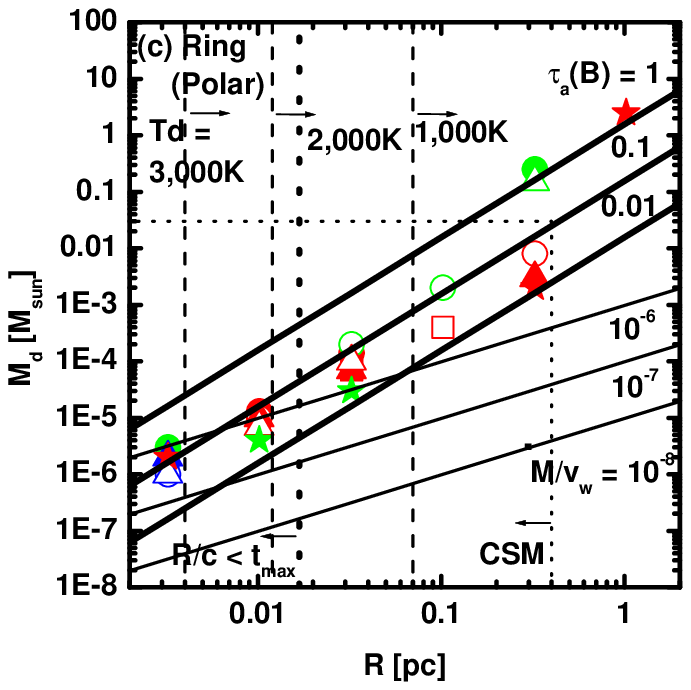}
 \includegraphics[width=0.33\textwidth]{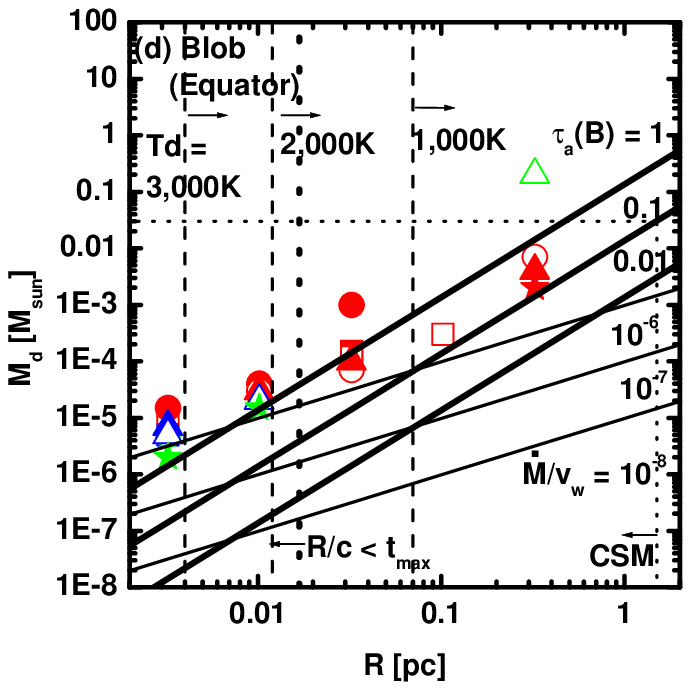}
 \includegraphics[width=0.33\textwidth]{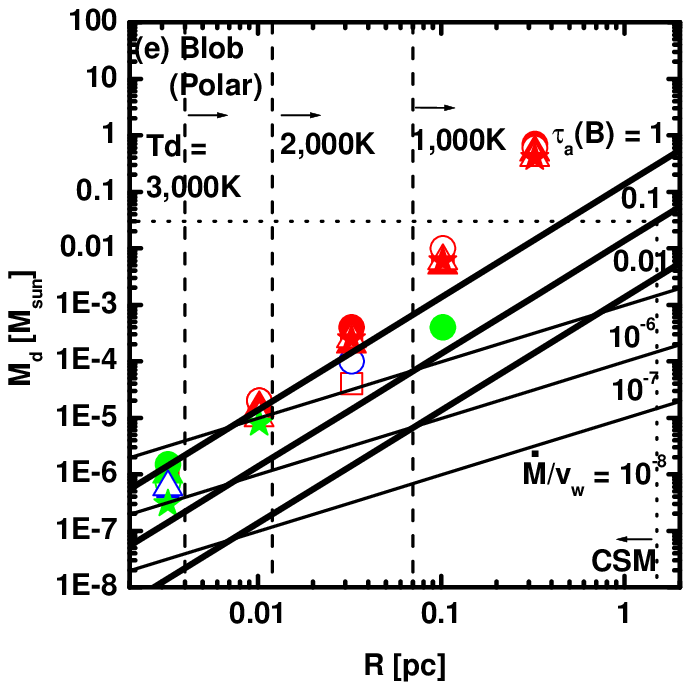}
\caption
{Conservative upper limit on the amount of CS dust as a function of $R$ (distance to the CS dust), for the shell CSM (a), the ring-like CSM as viewed from the equator (b) and from the polar direction (c), and the bipolar blob-like CSM as viewed from the equator (d) and from the polar direction (e). The symbols indicate the observational sample for which the constraints are derived: CSP sample \citep[filled square:][]{folatelli2010}, SN 2006X \citep[open square:][]{wood2008}, SN 1998bu \citep[filled circle:][]{krisciunas2003}, SN 2000cx \citep[open circle:][]{sollerman2004}, SN 2001el \citep[filled triangle:][]{stritzinger2007}, SN 2003du \citep[open triangles:][]{motohara2006}, and SN 2003hv \citep[filled star:][]{motohara2006,leloudas2009}. For each radius and given SN, the symbol is coloured according to the photometric band that gives the tightest constraint: $J$ (blue), $H$ (green), and $K_{\rm s}$ (red). The thick solid lines indicate $M_{\rm d}$ corresponding to $\tau_{\rm a} (B) = 1$, $0.1$, and $0.01$. If the origin of the dust is `CS', $M_{\rm d}$ would not exceed a few $0.01 M_{\odot}$, corresponding to a few $M_{\odot}$ in the CSM mass (dotted horizontal line for $M_{\rm d} = 0.03 M_{\odot}$). For the CS dust to be the origin of the peculiar extinction law, $\tau_{\rm a} (B)$ should exceed at least $0.1$ (conservatively) -- the corresponding radius is indicated by the vertical dotted line, left of which is considered to be CSM and right as ISM. The dashed vertical lines indicate the radius at which the equilibrium temperature of the CS dust is 3,000 K, 2,000 K, or 1,000 K. Assuming a gas-to-dust ratio of $100$, the equivalent steady-state mass loss rate is shown by the thin solid lines (in unit of $M_{\odot} / 10 \ {\rm km} \ {\rm s}^{-1}$). }
\label{fig4}
\end{figure*}

By applying different models (i.e., with a different $R$ and geometry) to the available, observed, multi-band light curves, a constraint can be obtained for $M_{\rm d}$ as a function of $R$ for each SN. The light curves for different bands provide mutually independent upper limits; we adopted the strongest limit among $J$, $H$, and $K_{\rm s}$. The result is shown in Fig. 4, where the method was applied to a `typical NIR observation' using the CSP sample, SN 2006X, as an example of very intensively observed early-phase light curves, including the $K_{\rm s}$-band and extending to a relatively late epoch; also shown are SNe 1998bu, 2000cx, 2001el, 2003du, and 2003hv, for which NIR late-phase ($\sim 1$ year) magnitudes are available. It is seen that generally the strongest constraints are obtained through the $J$ or $H$-band data for the CS dust at $R \lsim 0.1$ pc, and through the $K_{\rm s}$-band at $R \gsim 0.1$ pc.

Figure 4 also contains lines indicating various physical situations. The thick solid lines indicate the absorptive optical depth in the $B$-band, which gives indication of the likelihood that the CS dust is the origin of the non-standard extinction law (i.e., an optical depth of an order unity is required). The thin solid lines show the equivalent mass loss rates of the progenitor systems; we adopted the dust-to-gas ratio of 0.01 (i.e., $\dot M / v_{\rm w} = 100 M_{\rm d}/R$, where $v_{\rm w}$ is the velocity of the mass loss outflow). Vertical dashed lines show the temperature of the CS dust as a function of $R$. 

Figure 4 shows that the upper limits for $M_{\rm d}$, as obtained in the argument using the NIR echo (with the shell model), are $M_{\rm d} \lsim 10^{-5} M_{\odot}$ for the CS dust at $R \lsim 0.01$ pc, $M_{\rm d} \lsim 10^{-4} M_{\odot}$ at $R \sim 0.03$ pc, and $M_{\rm d} \lsim 10^{-2} M_{\odot}$ at $R \sim 0.1$ pc. Beyond this radius, the results are especially sensitive to the quality of the observational data, as essentially one needs still-rare late-time photometry for the $K_{\rm s}$-band. The strongest constraints for $R \gsim 0.1$ pc are obtained for SNe 2000cx, 2001el, and 2003hv, which are only three examples to date having the late time $K_{\rm s}$-band photometry at $\sim 1$ year. For these SNe, $M_{\rm d} \lsim 0.01 M_{\odot}$ is derived for $R \sim 0.4$ pc. At $R \gsim 1$ pc, even the NIR bands are too blue; thus, we can only provide a very loose constraint on the CS dust mass as $M_{\rm d} \lsim 1 M_{\odot}$, even for the best-case scenario (SN 2003hv). To probe the CS dust beyond $R \sim 1$ pc, mid- or far-IR observations should be more effective \citep{johansson2013,johansson2014}. 

\begin{figure*}
\centering
 \includegraphics[width=0.33\textwidth]{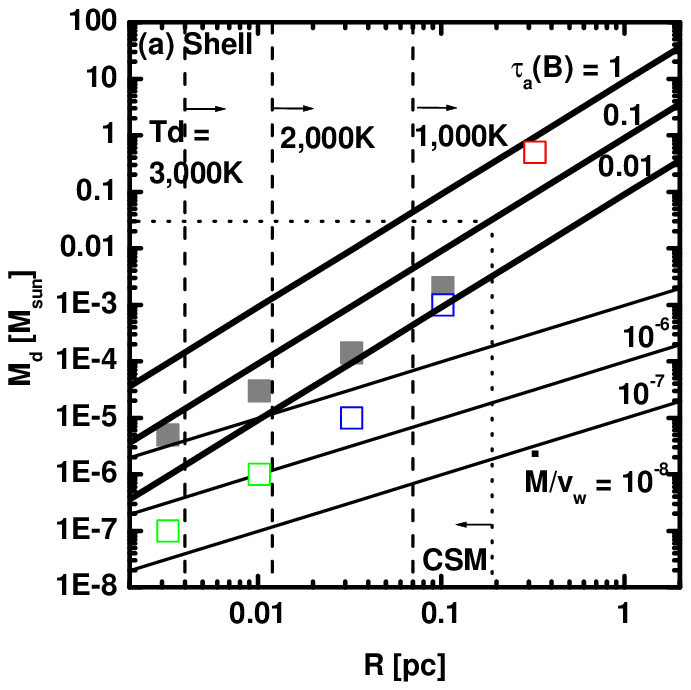}
 \includegraphics[width=0.33\textwidth]{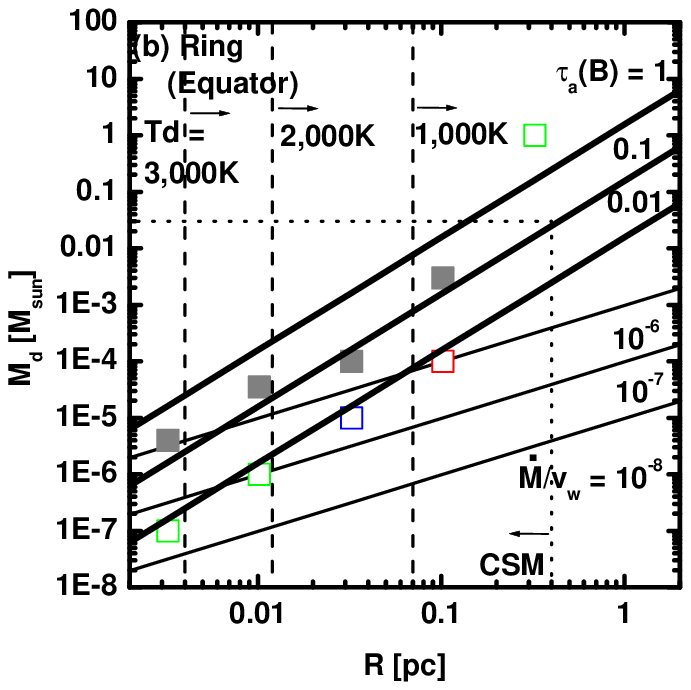}
 \includegraphics[width=0.33\textwidth]{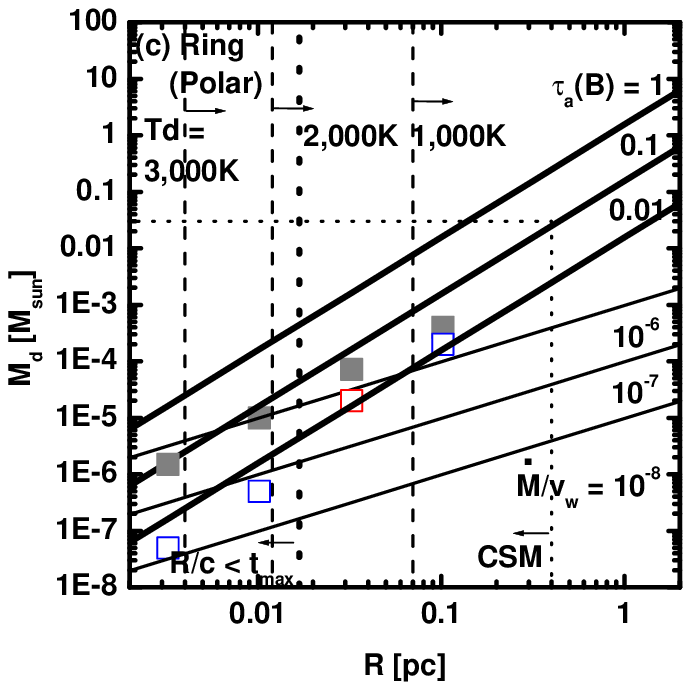}
 \includegraphics[width=0.33\textwidth]{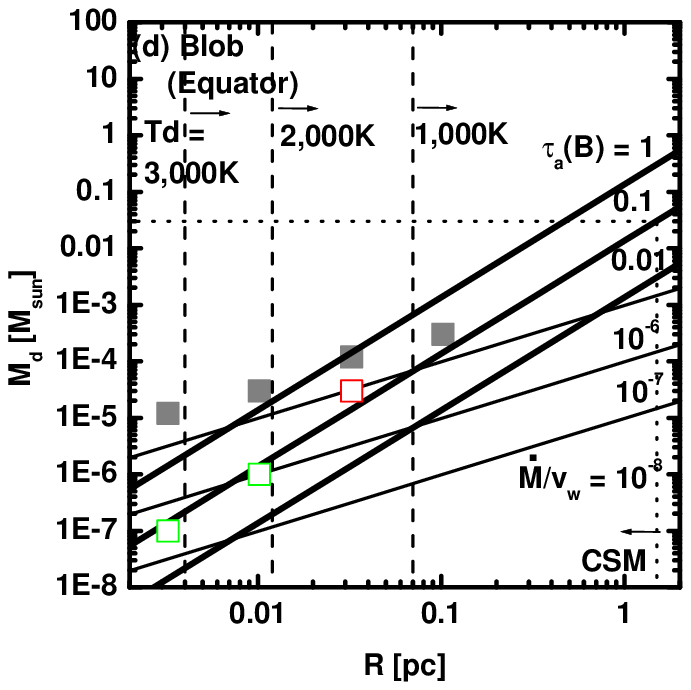}
 \includegraphics[width=0.33\textwidth]{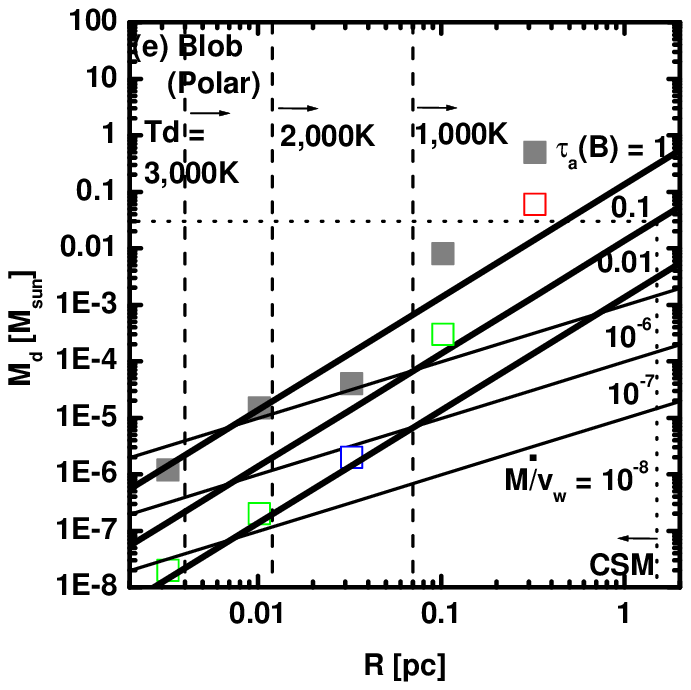}
\caption
{Deeper upper limit on the amount of CS dust as a function of $R$ (distance to the CS dust), for the shell CSM (a), the ring-like CSM viewed from the equator (b) and from the polar direction (c), and the bipolar blob-like CSM as viewed from the equator (d) and from the polar direction (e). Here, an upper limit is placed on the requirement that the resulting echo flux must be below the difference between the observed magnitude and the template magnitude. The deeper limit is shown by the open squares (blue, green, and red constrained in $J$, $H$, and $K_{\rm s}$, respectively); the conservative limit is shown as filled grey squares to show the potential for improvement in the constraint.} 
\label{fig5}
\end{figure*}

The derived upper limits are sufficiently tight to constrain the origin of the non-standard extinction as a scattered echo, as proposed by \citet{wang2005} and \citet{goobar2008}. The resulting upper limit for the optical depth in the $B$-band is generally $\lsim 0.1$ for the CS dust at $R \lsim 0.1$ pc. At $R \gsim 0.1$ pc, the constraint is weaker, but for a few SNe, we place the constraint as $\tau_{\rm a} (B) \lsim 0.01$ at $R \sim 0.4$ pc. Above $R \sim 1$ pc, the upper limit is consistent with $\tau_{\rm a} (B) \sim 1$. These values should be compared to the requirement that the optical depth must be at an order of unity for the scattered echo to have a substantial effect on the extinction law. From this analysis, we conclude that it is difficult for the scattered echo to be a general origin of the non-standard extinction law toward SNe, as long as the CS dust is placed at $R \lsim 0.1$ pc. A possibility that the CS dust at $R \gsim 1$ pc creates the non-standard extinction law is not rejected,  although categorising it as `CS' dust is questionable (see below). A dust shell at $R \sim 0.1 - 1$ pc is rejected for some SNe Ia when the late-time $K_{\rm s}$-band data are available, but not generally constrained due to a lack of such data. 

The above constraints are for the shell model. One might guess that the constraint would be weakened if the CS dust is confined in a specific direction, e.g., in an equatorial ring or in bipolar blobs, as the CS dust mass required to create $\tau_{\rm a} (B) \sim 1$ should be reduced. While this is true, several additional effects must be considered. First, the shell model does indeed provide the most conservative upper limit on the CS dust mass. If one confines the same amount of CS dust in a specific direction, then the local density increases. This generally results in larger luminosity at a specific epoch than in the corresponding shell model, and thus a tighter limit on $M_{\rm d}$ is obtained using the observational data at the corresponding epoch (Fig. 2). For example, the edge-on ring model predicts a larger luminosity than the shell model with the same CS dust mass at the beginning and the end of the NIR echo. In contrast, the same model viewed pole-on has a shorter duration and a larger luminosity. This large luminosity in the ring model is achieved when the emission at the ring-edge reaches the observer. A similar argument applies to the blob model.

As for the required CS dust mass within the scattering echo model for the extinction law, it is true that the required mass is reduced for an edge-on observer to the ring-like CS dust structure (Fig. 4). However, for the pole-on case, this effect is compensated for by an increase in the predicted luminosity (Fig. 2). Thus, the resulting upper limit for $\tau_{\rm a} (B)$ in the pole-on ring model is similar to that of the shell model (Fig. 4). Statistically, there must be edge-on and pole-on counterparts; the NIR echo constraint in this paper, using a number of SNe, is sufficiently strong to reject this possibility. If one introduces a {\em more confined} configuration, then this increases the likelihood that it will not be viewed from that direction. In addition, for a pole-on observer to have a substantial scattering effect, the time required for light to travel to the CS dust must be shorter than the rising time of SNe Ia (i.e., $\sim 15 - 20$ days), and therefore the CS dust beyond $R \sim 0.02$ pc never affects the extinction law if viewed pole-on. 

The above argument should be slightly modified for a very confined case, such as when considering the bipolar blob geometry. Similar to the ring-like geometry, the upper limits on the CS dust mass for a bipolar blob are comparable to or deeper than the corresponding shell model, depending on the viewing direction. If this is converted into the optical depth, the derived upper limits are indeed not extremely strong to readily reject the multiple scattering model for the extinction law, i.e., $\tau_{\rm a} (B) \sim 1$ (Fig. 4). However, in this case, the amount of scattered photons should already be limited, due to a small opening angle subtended by the distribution of CS dust. In summary, introducing the confined CS geometry does not help interpret the non-standard extinction law as {\em generally} having come from CS scattered light. SNe suffering a substantial `CS' scattering echo effect must be limited to at most only a fraction of SN Ia samples. 

For the CS dust to have originated from the mass loss of a progenitor, the CS gas mass should be $\lsim$ a few $M_{\odot}$, i.e., $M_{\rm d} \lsim$ a few $\times 10^{-2} M_{\odot}$. This amount of CS dust can create a substantial scattering effect, only if $R \lsim 0.2$ pc, even if we use the conservative requirement $\tau_{\rm a} (B) \sim 0.1$ (or $\sim 0.05$ pc if we adopt $\tau_{\rm a} (B) \sim 1$). We generally reject such a large amount of CS dust at $R \lsim 0.2$ pc (see above). Therefore, if the echo scenario is to be a general origin of the non-standard extinction law, then the dust must indeed be ISM in origin, located at $R \gsim 0.2$ pc. 

One may question what implications this mass loss rate may have for the progenitor scenario. Indeed, the CS scattering model requires a much larger amount of CSM, compared to that predicted by typical progenitor scenarios. Our limit is $\dot M \lsim 10^{-6} M_{\odot}$ yr$^{-1}$ for the mass-loss wind velocity of $\sim 10$ km s$^{-1}$ below 0.01 pc, or, in the final $\sim 400$ years before the explosion. This is not as strong as the constraint from radio and X-rays \citep[e.g., $\dot M \lsim 10^{-9} M_{\odot}$ yr$^{-1}$ for the same velocity, for SN 2011fe:][]{chomiuk2012,margutti2012}. However, we emphasise that radio and X-rays probe the materials inside $R \sim 3 \times 10^{-3}$ pc; our proposed method using the NIR echo can extend the constraint by an order of magnitude in the spatial dimension (or to the past before the explosion), making our limit independent and unique. From the present analysis (see below for possible improvement), the only scenario that can be marginally rejected, for {\em normal} SNe examined in this paper, is a symbiotic system. However, we do note that SNe Ia resulting from this path may appear as an atypical SN Ia \citep{dilday2012}. 

Thus far, our limit has been rather conservative, and is based on the condition that the echo luminosity cannot exceed the observed one. However, most of the light should have originated from a radioactive decay chain and subsequent thermalisation, as demonstrated by NIR spectroscopy \citep{gerardy2007,marion2009,hsiao2013}. Given this observational fact, we can refine our limit as follows. First, attribute the {\em template} light curves in each band to the SN light without any contribution of the possible NIR echo; subtract the templates from the light curves from individual SNe; and then use the residual light curve (here we take the absolute value if the subtraction results in a negative value) to place an upper limit on the CS dust mass. Namely, we assume that the echo luminosity does not exceed the {\em variation} in the NIR light curves, seen in different SNe. As a demonstration, in Fig. 3 (left panel) we show the case of SN 2006X, for which a good light curve is available. The figure shows that the magnitude of this residual light curve is $1 - 5$ less than that of the original, highlighting the standard nature of the NIR light curves. 

Using this argument, much tighter constraints can be obtained (as shown by the thin solid lines in Fig. 3). Figure 5 shows the results for SN 2006X. Because $J$- and $H$-band data are of the best quality, following the template closely, the upper limit for the CS dust mass, especially at $R \lsim 0.1$ pc, improved significantly. For small $R$, the upper limit was reduced by two orders of magnitude, which reflects a residual light curve that was fainter than the original by $\sim 5$ magnitudes. With this method, all of the models with different geometric types result in $\tau_{\rm a} (B) \lsim 0.1$ ($\tau_{\rm a} \lsim 0.01$ for most situations), robustly rejecting the CS scattering model for the extinction law for this particular SN. The upper limit for the mass loss rate is now $\dot M \lsim 10^{-7} M_{\odot}$ yr$^{-1}$ for $R \lsim 0.01$ pc, much deeper than the conservative limit, as described earlier in this section. 

\section{Discussion}
\subsection{Dust properties}
In this paper, our analysis has been based on a particular dust model, i.e., the LMC-type dust opacity. In this section, we examine the effects of different types of CS dust. Figure 6 shows the `conservative' limits on the CS dust mass derived for (a) astronomical silicate and (b) amorphous carbon or graphite, both for the shell model. Figure 7 shows the same but for the `deeper limit' for SN 2006X (i.e., Figure 5 but for astronomical silicate or carbon). The opacities are calculated using optical constants by \citet{draine2003} and \citet{zubko1996}. To the first approximation, the behaviour can be understood by the difference in a typical opacity for different dust models, e.g., the opacity in the $B$-band. The opacity is roughly a factor of two lower for silicate than the reference LMC-like dust model; thus, the upper limit for the CS dust mass is weakened by the same factor. The opacity is roughly an order of magnitude higher for carbon than the LMC-like dust model, and therefore the upper limit on the CS dust mass is deeper for the carbon dust model by an order of magnitude than our fiducial model. 

However, these different constraints on the CS dust mass would not substantially alter the constraints on the ability of the CS dust to create the non-standard extinction law. The optical depth itself is scaled by the same factor; therefore, the upper limits on $\tau_{\rm a} (B)$ would not change substantially.

\begin{figure*}
\centering
 \includegraphics[width=0.33\textwidth]{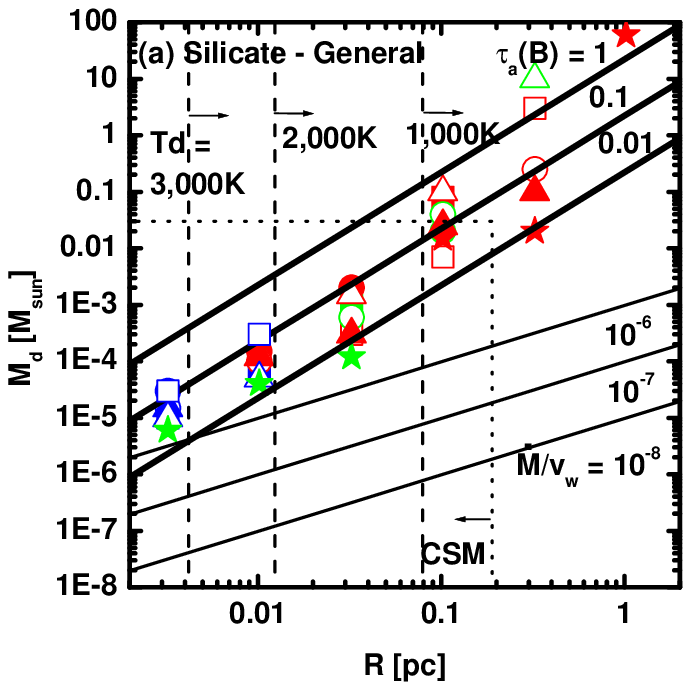}
 \includegraphics[width=0.33\textwidth]{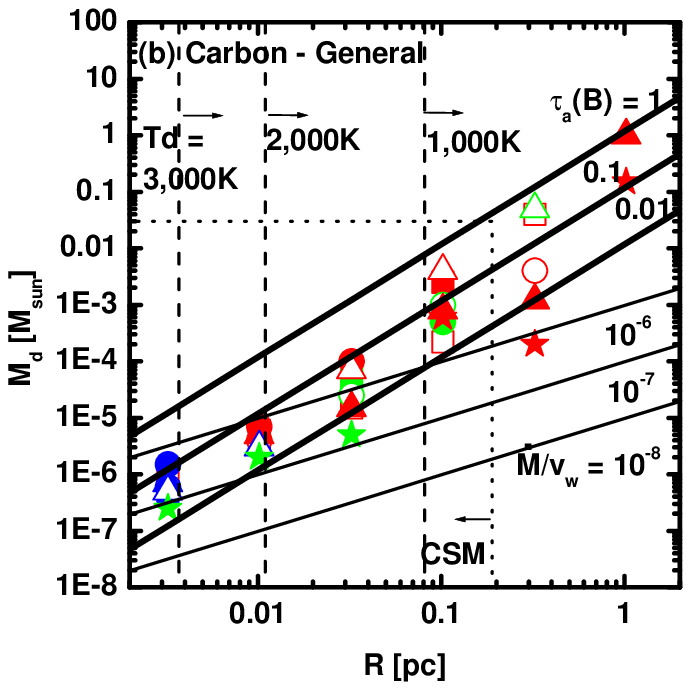}
\caption
{Same as Fig. 4, but (a) for astronomical silicate dust and (b) for amorphous carbon/graphite. 
}
\label{fig6}
\end{figure*}

\begin{figure*}
\centering
 \includegraphics[width=0.33\textwidth]{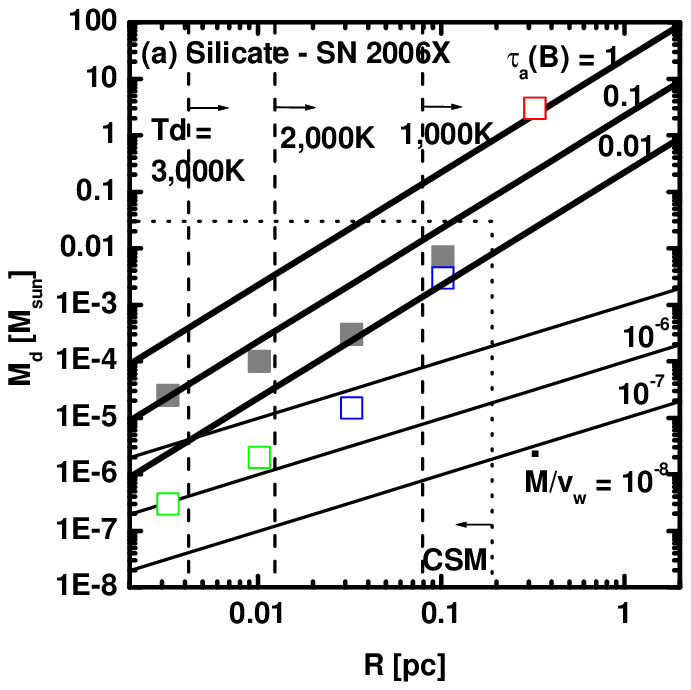}
 \includegraphics[width=0.33\textwidth]{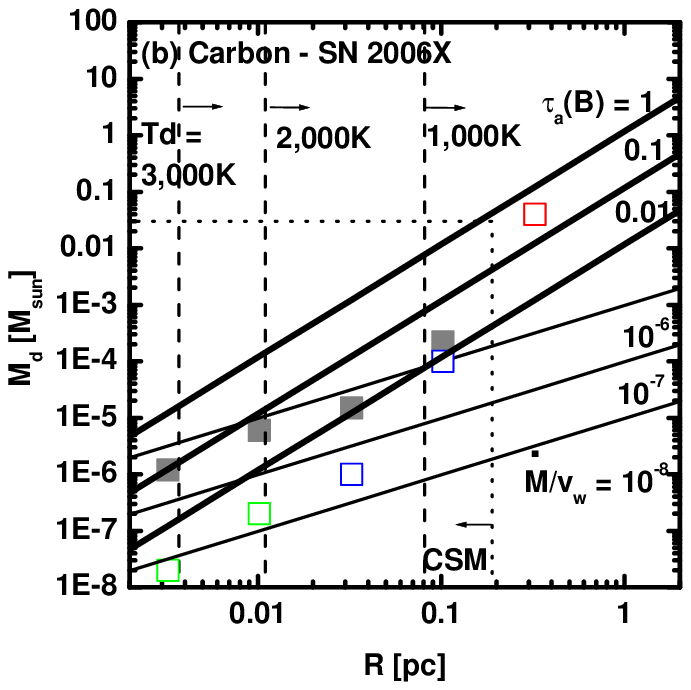}
\caption
{Same as Fig. 5, but (a) for astronomical silicate dust and (b) for amorphous carbon/graphite.
}
\label{fig7}
\end{figure*}

\subsection{Implications for some specific cases} 
Our results should be compared to SNe, in which a large amount of dust is inferred either by the detection of an optical (scattered) echo or other probes (e.g., narrow absorption systems). For example, we have obtained a tight limit for the amount of CS dust for SN 2006X (Fig. 5), while this SN was detected in a scattered echo \citep{wang2008}; its optical spectrum at $\sim 1$ year after the explosion showed a blue echo component, indicating $\tau_{\rm a} (B) \sim 1$. In contrast, we obtained the constraint that $\tau_{\rm a} (B) < 1$, as long as the echo source is at $R \lsim 0.4 $pc (or even $\tau_{\rm a} \lsim 0.01$ at $R \lsim 0.1$ pc). Taken together, the detected echo in SN 2006X must be created by ISM, as suggested by \citet{crotts2008} using Hubble Space Telescope ({\em HST}) data. To date, even for cases where an (optical) echo is detected, evidence of a CS echo remains weak. 

\begin{figure*}
\centering
 \includegraphics[width=0.7\textwidth]{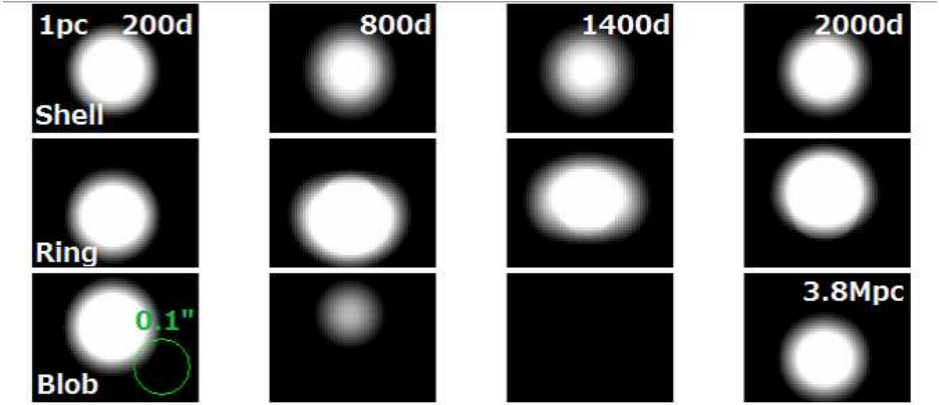}
\caption
{Simulated images of a CS echo in the $K_{\rm s}$-band, for an SN at a distance of $3.8$ Mpc, observed with a spatial resolution of $0.1$ arcsec. The distance to the CS dust is set as $R = 1$ pc. The images are created by smoothing the echo model by a Gaussian kernel with $0.1$ arcsec (shown by a green circle). The time evolution is shown from the left to the right, and different geometric models are shown from top to bottom. An SN at this distance is represented by SN 2014J. The adopted resolution slightly overestimates the {\em HST} resolution in the $K$-band ($\sim 0.15$ arcsec), but here it is adopted for the purpose of demonstration; since the angular size scales linearly with the physical scale, this relative-flux map applies to the case with $R = 1.5$ pc if the angular resolution is worse by $50$\%.
}
\label{fig8}
\end{figure*}

Another example is highly extinct SN Ia 2014J in M82. No variability has been found in the narrow absorption lines \citep[e.g., ][]{goobar2014}. \citet{foley2014} have claimed temporal evolution in $R_{V}$, while no apparent variability has been detected in polarisation measurements \citep{kawabata2014}. While a consistent interpretation for all of these features has yet to be reached, it seems that together they suggest that the CS scattering scenario is unlikely, but perhaps a combination of multiple scattering and true absorption at the ISM scale could provide a consistent solution. Indeed, detection of an echo has been reported for SN 2014J through high-resolution {\em HST} images at $\sim 230$ days after the explosion \citep{crotts2014}. This has been interpreted as echoes originating at the interstellar scale. At this epoch, the SN still dominates the light in the vicinity of the SN and no strong signature of the echo from the CS dust has been obtained \citep[see also][]{johansson2014}.

In Fig. 8, we show the results of simulated $K_{\rm s}$-band images of an echo from the CS dust located at $R = 1$ pc for different geometries, for an SN at $3.8$ Mpc. For the ring and blob geometries, we have adopted $\theta_0 = 30^{\circ}$ and the viewing direction of $45^{\circ}$ for the purpose of demonstration. The model images are convolved with a Gaussian kernel of $0.1$ arcsec to roughly simulate observations by {\em HST}; the Narrow field channel of {\em HST/NICMOS} has a spatial resolution of $\sim 0.14$ arcsec at $\sim 1.6 \micron$. Thus our simulation could overestimate the spatial resolution up to $\sim 50$\%, but it could anyway be compensated by increasing the assumed physical scale of the CS dust by the same factor. Note that these images are for a `pure' echo component -- if the projected echo size is smaller than or comparable to the instrument's spatial resolution, then the SN must fade away below the level of the echo luminosity to examine the geometry of the echo signal. Note that there is a better chance to apply this exercise to later epochs. Figure 8 shows the expected evolution covering $\sim 5$ years, to investigate the possibility of distinguishing different CS dust geometries if a CS dust echo is detected. Even for nearby SN 2014J and using {\em HST}, probing the CS dust echo using spatially resolved imagery can be difficult. 

For a ring-like structure, an elongated structure is visible when $t \sim R/c$. For this particular bipolar blob model, the echo `disappears' when $t \sim R/c$. For both models, there is a (non-monotonic) shift in the centre of the echo light. The size of this shift for these particular models is about half of the spatial resolution, and could be detectable with precision astrometry techniques. These diagnostics could apply to SN 2014J if the CS dust echo is detected in the future following the SN luminosity decrease. Continuous {\em HST} high-resolution observations are strongly encouraged. 

\subsection{Future possibility of high-resolution imaging observations}
While investigation of the CS echo through direct imaging observations is currently limited by the spatial resolution of $\sim 0.1$ arcsec at NIR wavelengths even with {\em HST} (\S 4.2), the situation is expected to improve dramatically in the coming decade. The planned James Webb Space Telescope ({\em JWST}) will reach to a spatial resolution by a factor of $\sim 2.5$ better than {\em HST}. Furthermore, with the availability of NIR adaptive optics (AO) attached to $30$ m class telescopes such as the Thirty Meter Telescope (TMT), European Extremely Large Telescope (E-ELT), and Giant Magellan Telescope (GMT), spatial resolutions of $\sim 0.01$ arcsec are expected. Figures 9 and 10 show synthesised images at different epochs for an SN at a distance of $3.8$ Mpc, surrounded by CS dust shell/ring/blobs, as viewed with a spatial resolution of $0.01$ arcsec (for the viewing direction of $45^{\circ}$ as a specific example). With this resolution, one can distinguish different CS dust geometries, even for the CS dust located at $R = 0.3$ pc. For a spherically symmetric distribution, investigation of the radial distribution is within reach, as seen by the hole for the shell model. The ring and blob models show asymmetric image profiles that are clearly different from the shell model (but depend on the viewing direction). The shift of the centre in these models should be easily resolved beyond the spatial resolution -- for example, using this shift, the blob model is distinguishable from any spherically symmetric distribution. 

An SN at a distance of $3.8$ Mpc is a rare event, at an observed rate of one per decade or two. Figures 11 and 12 show the same images, but for an SN at a distance of $15$ Mpc. At this distance, we expect an event rate of roughly one per year; thus, such observations are, in principle, annual occurrences. For CS dust located at $R = 0.3$ pc around an SN at a distance of $15$ Mpc, as observed by {\em TMT/E-ELT/GMT} (Fig. 11), the situation is similar to the case shown in Fig. 8. An elongated profile of the ring distribution could be marginally resolved, and the shift in images for asymmetrically distributed CS dust (e.g., ring or blobs) could be detectable by a sub-spatial resolution astrometry. For the CS dust at $R = 1$ pc, observations with a spatial resolution of $0.01$ arcsec can distinguish various CS dust geometries, given that such echo signals are detected. 

Thus, we conclude that the proposed future high-resolution NIR AO can open up a new window to examine the CS environment around nearby SNe. We note that a target does not have to be one discovered after the new instruments become available. For example, SN 2014J can already be a target -- assuming an observation at year 2024, CS dust located at $\gsim 2$ pc can still be investigated. Pilot work on this measurement may already be possible with $8$ m class telescopes approaching the diffraction limit ($\sim 0.05$ arcsec). 

\begin{figure*}
\centering
 \includegraphics[width=0.7\textwidth]{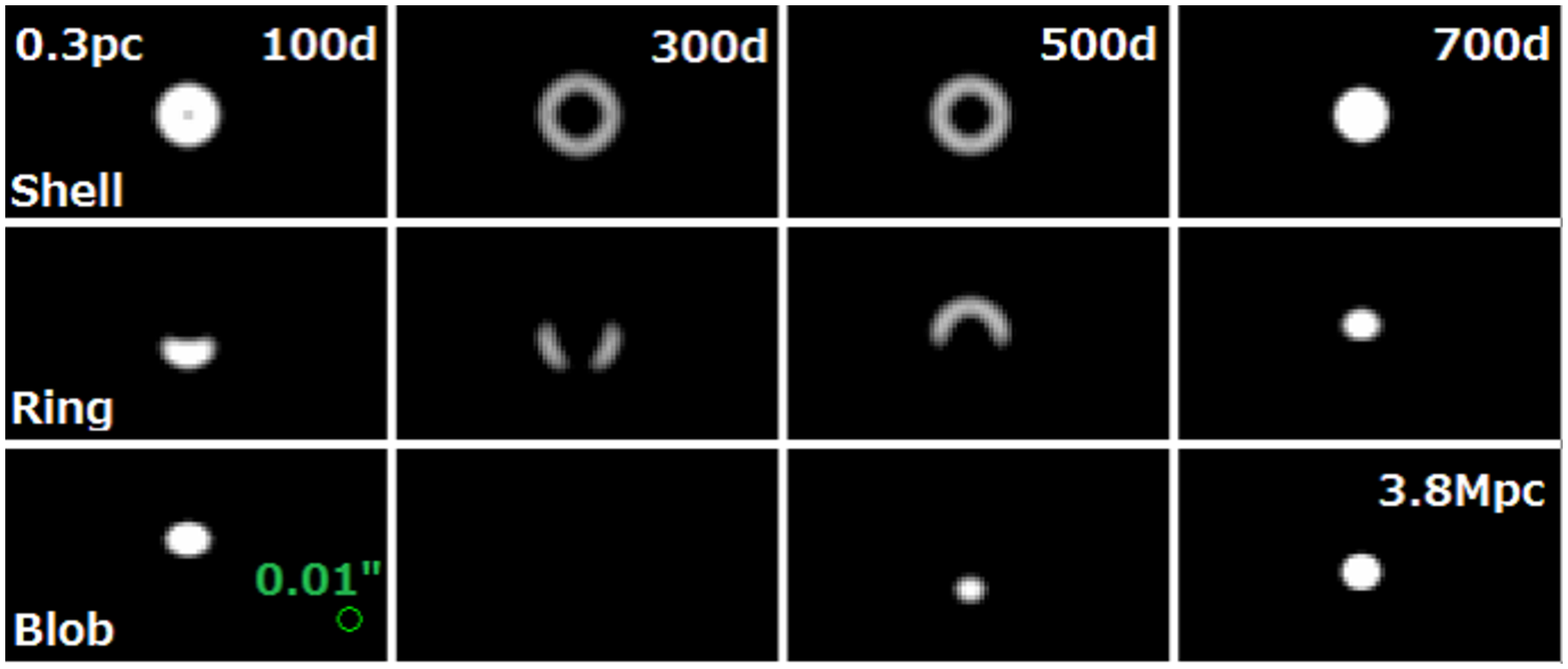}
\caption
{Simulated images of a CS echo in the $K_{\rm s}$-band, for an SN at a distance of $3.8$ Mpc as observed with a spatial resolution of $0.01$ arcsec (i.e., {\em TMT,E-ELT, GMT}). The distance to the CS dust is set as $R = 0.3$ pc. The images are created by smoothing the echo model by a Gaussian kernel with $0.01$ arcsec (shown by a green circle). The time evolution is shown from the left to the right, and different geometric models are shown from top to bottom. An SN at this distance is represented by SN 2014J. 
}
\label{fig9}
\end{figure*}

\begin{figure*}
\centering
 \includegraphics[width=0.7\textwidth]{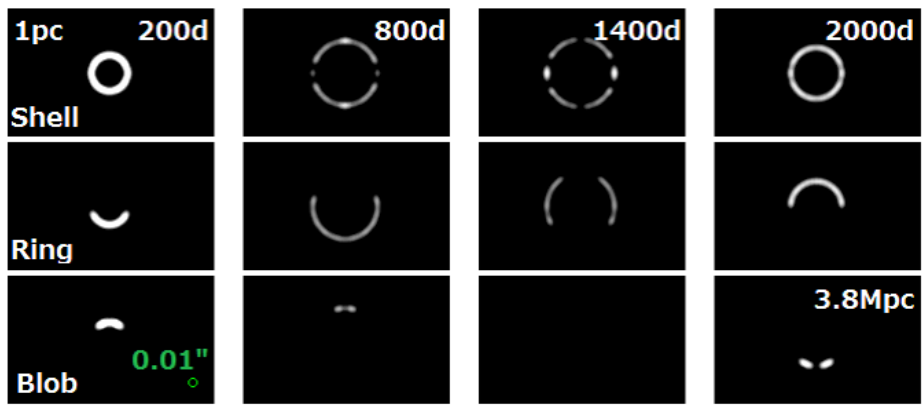}
\caption
{Same as Fig. 9, except that $R = 1$ pc. 
}
\label{fig10}
\end{figure*}

\begin{figure*}
\centering
 \includegraphics[width=0.7\textwidth]{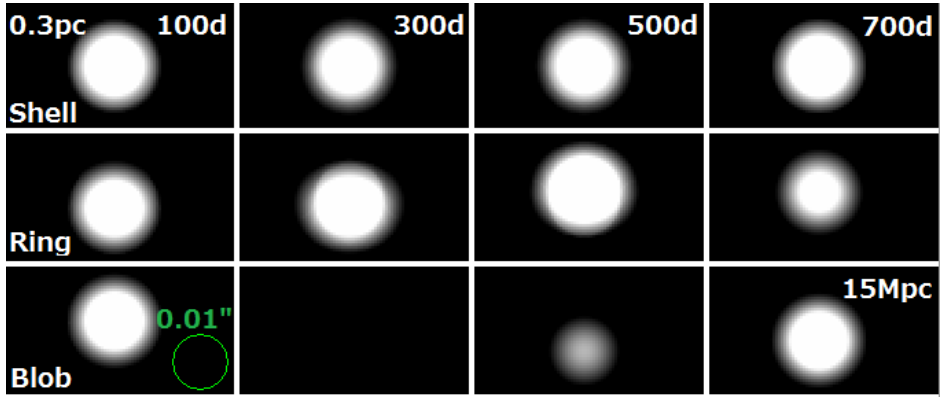}
\caption
{Same as Fig. 9, but for an SN at a distance of $15$ Mpc. On average, one SN per year has been discovered within $15$ Mpc. 
}
\label{fig9}
\end{figure*}

\begin{figure*}
\centering
 \includegraphics[width=0.7\textwidth]{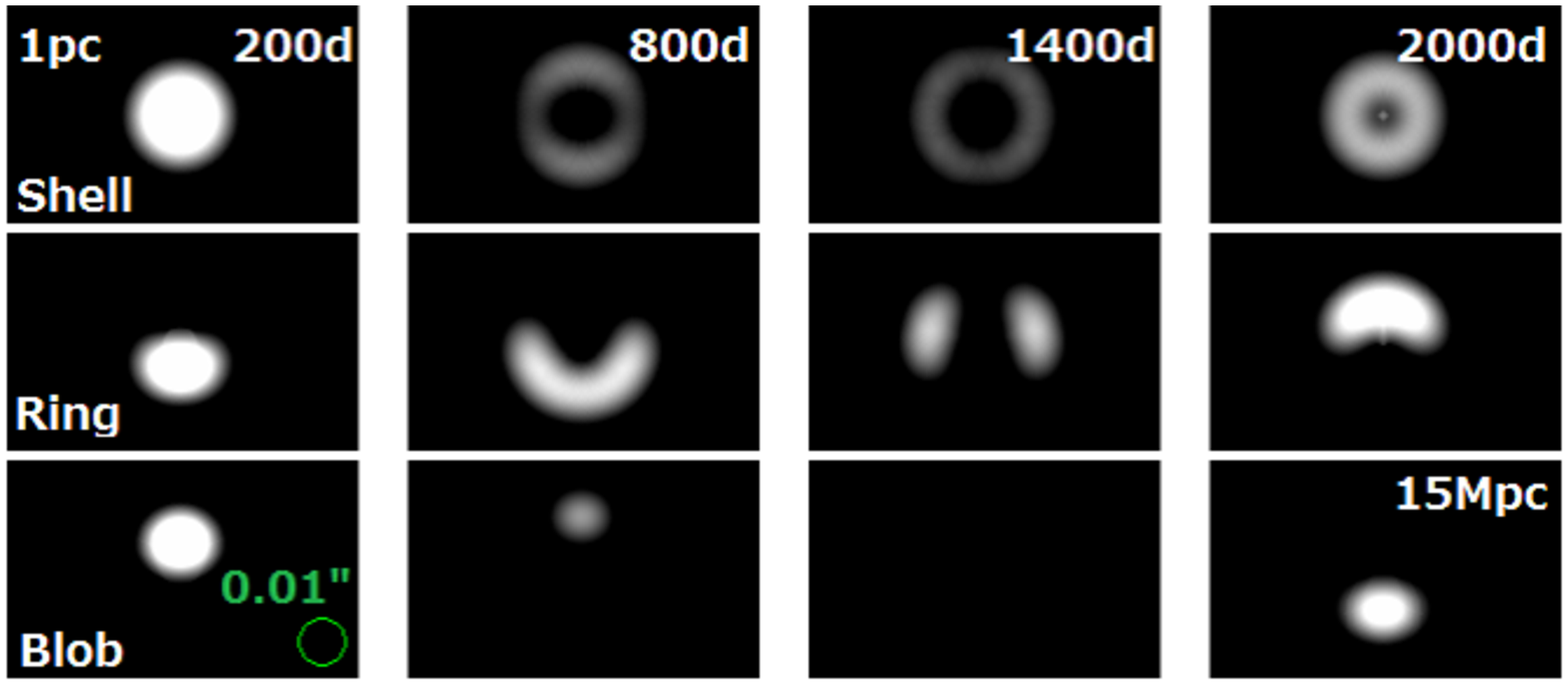}
\caption
{Same as Fig. 11, except that $R = 1.0$ pc. 
}
\label{fig10}
\end{figure*}

\section{Conclusions}
We investigated the effects of {\em re-emission} of SN photons by CS dust for IR wavelengths. We have shown that this effect allows observed IR light curves to be used to place constraints on the position/size and the amount of CS dust. We applied the method to observed NIR SN Ia samples, and found that meaningful upper limits on the CS dust mass can be derived, even under conservative assumptions.

Our results are summarised as follows: 
\begin{enumerate}
\item $M_{\rm d} \lsim 10^{-5} M_{\odot}$ for the CS dust at $R \lsim 0.01$ pc, $M_{\rm d} \lsim 10^{-4} M_{\odot}$ at $R \sim 0.03$ pc, and $M_{\rm d} \lsim 10^{-2} M_{\odot}$ at $R \sim 0.1$ pc. 
\item The mass loss rate must be generally $\dot M \lsim 10^{-6} M_{\odot}$ yr$^{-1}$ up to $R \sim 0.04$ pc, rejecting a symbiotic path as a main population of the {\em normal} SNe Ia. 
\item For a particular case of SN 2006X, a deeper limit is obtained (see below): $M_{\rm d} \lsim 10^{-6} M_{\odot}$ for the CS dust at $R \lsim 0.01$ pc, $M_{\rm d} \lsim 10^{-5} M_{\odot}$ at $R \sim 0.03$ pc, and $M_{\rm d} \lsim 10^{-3} M_{\odot}$ at $R \sim 0.1$ pc. The corresponding mass loss rate (upper limit) is $\dot M \lsim 3 \times 10^{-8} M_{\odot}$ yr$^{-1}$ up to $R \sim 0.004$ pc and $\dot M \lsim 10^{-7} M_{\odot}$ yr$^{-1}$ up to $R \sim 0.01$ pc.
\item The above upper limits become tighter if one considers a non-spherical CS distribution. 
\item The above upper limits become even tighter by an order of magnitude for carbon dust.
\end{enumerate}

From these results, we conclude the following regarding the possibility of the dust-scattering scenario for the non-standard extinction law toward SNe Ia: 
\begin{enumerate}
\item The `CS' dust (at $R \lsim 1$ pc) echo cannot be a general explanation of the non-standard extinction law with regard to SNe Ia. The scenario would work for at most a fraction of SNe Ia, presumably at most, the highly reddened ones. 
\item If the echo scenario is responsible for the extinction law in general, then the dust must be located at $R \gsim 1$ pc. The required mass is, however, $\gsim 100 M_{\odot}$; therefore, this must be attributed to ISM rather than CSM. 
\end{enumerate}

We have also demonstrated an approach that allows deeper limits to be applied. This involves first using the standard nature of the NIR light curves. We demonstrated this for SN 2006X, showing that the upper limit can be deepened by two orders of magnitude from the conservative limit, as given above. We suggest that further improvement is possible by refining the analysis method along this line via standardisation of the NIR light curves for a given sample and minimisation of the dispersion as a function of time. This would allow a deeper limit to be placed on the mass loss rate, potentially to the level to limit various models other than the particular symbiotic scenario. Again, one of the main advantages of the proposed method, compared to radio and X-ray constraints (which are indeed complementary to the proposed method here), is the following: the limit to the amount of dust placed by the NIR echo extends an order of magnitude larger in the spatial dimension (or an order of magnitude to the past in the time dimension), reflecting the fact that the light speed (for the echo) is by an order of magnitude faster than the shock velocity (for radio and X-rays). 

As we have shown in this paper, late-time photometry, even sparse, especially in the $K_{\rm s}$-band, is essential to constrain dust (and gas) mass in the environment at $R \gsim 0.1$ pc. Our method underlines the importance and usefulness of such data. Our analysis is also highlighted by the possibilities of (1) constraining the origin of the non-standard extinction law, and (2) constraining the CSM mass up to $\sim 1$ pc, namely the regimes that the other methods like radio and X-ray diagnostics cannot probe. Because the data are still rare, we propose increasing the sample of NIR/mid-IR SNe with continuous observational efforts. Of course, if one wants to go further toward the ISM regime, mid- and far-IR data are important \citep{johansson2013,johansson2014}, and a combination of the early radio/X, early and late-time NIR, and late-time mid-/far-IR, perhaps complemented by the absorption system study \citep[e.g.,][]{phillips2013}, will hopefully provide a complete view of these issues. 

We also investigated the possibility of studying the CS dust distribution through high spatial resolution observations. Our results are summarised below. 
\begin{enumerate}
\item CS dust around SN 2014J could potentially be resolved by future {\em HST} observations, given that dense CS dust shell/ring/blobs exist at $R \gsim 1$ pc from the SN site. 
\item Future NIR AO instruments attached to $30$ m class telescopes are capable of distinguishing various distributions of CS dust, thus providing a powerful means for investigating the CS environment around SNe Ia. 
\item Such observations will be possible for nearby SNe at a rate of roughly once a year. 
\end{enumerate}

\section*{Acknowledgments}
The authors thank Ariel Goobar for fruitful discussion. K.M. acknowledges financial support by Grant-in-Aid for Scientific Research (No. 23740141 and 26800100) from the Japanese Ministry of Education, Culture, Sports, Science and Technology (MEXT). The work by K.M. is partly supported by World Premier International Research Center Initiative (WPI Initiative), MEXT, Japan. The authors also thank the anonymous referee for her/his many constructive comments.

%\appendix

\bsp

\label{lastpage}


\begin{thebibliography}{99}

\bibitem[Amanullah \& Goobar(2011)]{amanullah2011}
Amanullah, R., \& Goobar, A., 2011, ApJ, 735, 20

\bibitem[Chesneau et al.(2012)]{chesneau2012} 
Chesneau, O., et al. 2012, A\&A, 545, A63

\bibitem[Chomiuk et al.(2012)]{chomiuk2012}
Chomiuk, L., et al. 2012, ApJ, 750, 164

\bibitem[Crotts \& Yourdon(2008)]{crotts2008}
Crotts, A., \& Yourdon, D. 2008, ApJ, 689, 1186

\bibitem[Crotts(2015)]{crotts2014}
Crotts, A.P.S. 2015, ApJ, 804, L37

\bibitem[Dilday et al.(2012)]{dilday2012}
Dilday, B., et al. 2012, Science, 337, 942 

\bibitem[Draine(2003)]{draine2003}
Draine, B.T. 2003, ApJ, 598, 1026

\bibitem[Dwek(1983)]{dwek1983}
Dwek, E., 1983, ApJ, 274, 175

\bibitem[Dwek(1985)]{dwek1985}
Dwek, E., 1985, ApJ, 297, 719

\bibitem[Dwek \& Felten(1989)]{dwek1989}
Dwek, E., Felten, J.E. 1989, ApJ, 342, 300

\bibitem[Felten \& Dwek(1989)]{felten1989}
Felten, J.E., Dwek, E. 1989, Nature, 339, 123

\bibitem[Folatelli et al.(2010)]{folatelli2010}
Folatelli G. et al., 2010, AJ, 139, 120

\bibitem[Foley et al.(2014)]{foley2014}
Foley, R., et al. 2014, MNRAS, 443, 2887

\bibitem[Gerardy et al.(2007)]{gerardy2007}
Gerardy, C.L., et al. 2007, ApJ, 661, 995

\bibitem[Goobar(2008)]{goobar2008}
Goobar, A., 2008, ApJ, 68, L103

\bibitem[Goobar et al.(2014)]{goobar2014}
Goobar, A., et al. 2014, ApJ, 784, L12 

\bibitem[Hachisu et al.(1999)]{hachisu1999}
Hachisu, I., Kato, M., Nomoto, K. 1999, ApJ, 522, 487

\bibitem[Hsiao et al.(2007)]{hsiao2007}
Hsiao, E. Y., Conley, A., Howell, D. A., Sullivan, M., Pritchet, C. J., Carlberg, R. G., Nugent, P. E., Phillips, M. M. 2007, ApJ, 663, 1187

\bibitem[Hsiao et al.(2013)]{hsiao2013}
Hsiao, E.Y., 2013, ApJ, 766, 72

\bibitem[Johansson et al.(2013)]{johansson2013}
Johansson, J., Amanullah, R., \& Goobar, A. 2013, MNRAS, 431, L43

\bibitem[Johansson et al.(2014)]{johansson2014}
Johansson, J., et al. 2014, arXiv:1411.3332

\bibitem[Kawabata et al.(2014)]{kawabata2014}
Kawabata, K.S., et al. 2014, ApJ, 795, L4

\bibitem[Leloudas et al.(2009)]{leloudas2009}
Leloudas, G., et al. 2009, A\&A, 505, 265

\bibitem[Krisciunas et al.(2003)]{krisciunas2003}
Krisciunas, K., et al. 2003, AJ, 125, 166

\bibitem[Margutti et al.(2012)]{margutti2012}
Margutti, R., et al. 2012, ApJ, 751, 134

\bibitem[Marion et al.(2009)]{marion2009}
Marion, G.H., et al. 2009, ApJ, 138, 727

\bibitem[Motohara et al.(2006)]{motohara2006}
Motohara, K., et al. 2006, ApJ, 652, L101

\bibitem[Nomoto(1982)]{nomoto1982}
Nomoto, K. 1982, ApJ, 253, 798

\bibitem[Phillips et al.(1999)]{phillips1999}
Phillips M.~M., Lira P., Suntzeff N.~B., Schommer R.~A., 
Hamuy M., Jose M., 1999, 
AJ, 118, 1766 

\bibitem[Phillips et al.(2013)]{phillips2013}
Phillips M.~M., et al. 2013, ApJ, 779, 38

\bibitem[Sollerman et al.(2004)]{sollerman2004}
Sollerman, J., et al. 2004, A\&A, 428, 555

\bibitem[Spyromilio et al. (2004)]{spyromilio2004}
Spyromilio, J., et al. 2004, A\&A, 426, 547

\bibitem[Stanishev et al.(2007)]{stanishev2007}
Stanishev, V., et al. 2007, A\&A, 469, 645

\bibitem[Stritzinger \& Sollerman(2007)]{stritzinger2007}
Stritzinger, M., Sollerman, J. 2007, A\&A, 470, L1

\bibitem[Wang(2005)]{wang2005}
Wang, L., 2005, ApJ, 635, L33

\bibitem[Wang et al.(2008)]{wang2008}
Wang, X., et al. ApJ, 677, 1060 

\bibitem[Whelan \& Iben(1973)]{whelan1973}
Whelan, J., Iben, Jr., I. 1973, ApJ, 186, 1007

\bibitem[Wood-Vasey et al.(2008)]{wood2008}
Wood-Vesey, W.M., et al. 2008, ApJ, 689, 377

\bibitem[Zubko et al.(1996)]{zubko1996}
Zubko, V.G., Mennella, V., Colangeli, L., Bussoletti, E. 1996, MNRAS, 282, 1321

\end{thebibliography}
\end{document}